\pdfoutput=1
\documentclass[12pt,a4paper]{article}
\usepackage{ifthen} % for conditional statements
\newboolean{pdflatex}
\setboolean{pdflatex}{true} % False for eps figures 

\newboolean{articletitles}
\setboolean{articletitles}{true} % False removes titles in references

\newboolean{uprightparticles}
\setboolean{uprightparticles}{false} %True for upright particle symbols

\usepackage{microtype}
\usepackage{lineno}  % for line numbering during review
\usepackage{xspace} % To avoid problems with missing or double spaces after
\usepackage{graphicx}  % to include figures (can also use other packages)
\usepackage{color}
\usepackage{colortbl}
\graphicspath{{./figs/}} % Make Latex search fig subdir for figures

\usepackage{amsmath} % Adds a large collection of math symbols
\usepackage{amssymb}
\usepackage{amsfonts}
\usepackage{upgreek} % Adds in support for greek letters in roman typeset

\newcommand*\patchAmsMathEnvironmentForLineno[1]{%
\expandafter\let\csname old#1\expandafter\endcsname\csname #1\endcsname
\expandafter\let\csname oldend#1\expandafter\endcsname\csname
end#1\endcsname
 \renewenvironment{#1}%
   {\linenomath\csname old#1\endcsname}%
   {\csname oldend#1\endcsname\endlinenomath}%
}
\newcommand*\patchBothAmsMathEnvironmentsForLineno[1]{%
  \patchAmsMathEnvironmentForLineno{#1}%
  \patchAmsMathEnvironmentForLineno{#1*}%
}
\AtBeginDocument{%
\patchBothAmsMathEnvironmentsForLineno{equation}%
\patchBothAmsMathEnvironmentsForLineno{align}%
\patchBothAmsMathEnvironmentsForLineno{flalign}%
\patchBothAmsMathEnvironmentsForLineno{alignat}%
\patchBothAmsMathEnvironmentsForLineno{gather}%
\patchBothAmsMathEnvironmentsForLineno{multline}%
}

\usepackage{hyperref}    % Hyperlinks in references
\usepackage[all]{hypcap} % Internal hyperlinks to floats.

\def\lhcb {\mbox{LHCb}\xspace}
\def\ux85 {\mbox{UX85}\xspace}

\ifthenelse{\boolean{uprightparticles}}%
{

 \def\Pmu         {\ensuremath{\upmu}\xspace}

 \def\Pphi        {\ensuremath{\upphi}\xspace}

 \def\Ppsi        {\ensuremath{\uppsi}\xspace}

 \def\PDelta      {\ensuremath{\Delta}\xspace}                 
 \def\PXi      {\ensuremath{\Xi}\xspace}                 
 \def\PLambda      {\ensuremath{\Lambda}\xspace}                 
 \def\PSigma      {\ensuremath{\Sigma}\xspace}                 
 \def\POmega      {\ensuremath{\Omega}\xspace}                 
 \def\PUpsilon      {\ensuremath{\Upsilon}\xspace}                 
 
 \def\PB      {\ensuremath{\mathrm{B}}\xspace}                 
                  
 \def\PD      {\ensuremath{\mathrm{D}}\xspace}

 \def\PJ      {\ensuremath{\mathrm{J}}\xspace}                 
 \def\PK      {\ensuremath{\mathrm{K}}\xspace}

 \def\Pb      {\ensuremath{\mathrm{b}}\xspace}

 \def\Pi      {\ensuremath{\mathrm{i}}\xspace}

}
{

 \def\Pmu         {\ensuremath{\mu}\xspace}

 \def\Pphi        {\ensuremath{\phi}\xspace}

 \def\Ppsi        {\ensuremath{\psi}\xspace}                 
                  
 \mathchardef\PDelta="7101
 \mathchardef\PXi="7104
 \mathchardef\PLambda="7103
 \mathchardef\PSigma="7106
 \mathchardef\POmega="710A
 \mathchardef\PUpsilon="7107
                  
 \def\PB      {\ensuremath{B}\xspace}                 
                  
 \def\PD      {\ensuremath{D}\xspace}

 \def\PJ      {\ensuremath{J}\xspace}                 
 \def\PK      {\ensuremath{K}\xspace}

 \def\Pb      {\ensuremath{b}\xspace}

 \def\Pi      {\ensuremath{i}\xspace}

}

\def\mup        {\ensuremath{\Pmu^+}\xspace}
\def\mun        {\ensuremath{\Pmu^-}\xspace} % muon negative (\mum is taken)

\def\bquark    {\ensuremath{\Pb}\xspace}

\def\kaon  {\ensuremath{\PK}\xspace}
  \def\Kbar  {\kern 0.2em\overline{\kern -0.2em \PK}{}\xspace}

\def\Kz    {\ensuremath{\kaon^0}\xspace}
\def\Kzb   {\ensuremath{\Kbar^0}\xspace}
\def\KzKzb {\ensuremath{\Kz \kern -0.16em \Kzb}\xspace}
\def\Kp    {\ensuremath{\kaon^+}\xspace}
\def\Km    {\ensuremath{\kaon^-}\xspace}

\def\KpKm  {\ensuremath{\Kp \kern -0.16em \Km}\xspace}
\def\KS    {\ensuremath{\kaon^0_{\rm\scriptscriptstyle S}}\xspace}

  \def\Dbar    {\kern 0.2em\overline{\kern -0.2em \PD}{}\xspace}
\def\D       {\ensuremath{\PD}\xspace}

\def\Dz      {\ensuremath{\D^0}\xspace}
\def\Dzb     {\ensuremath{\Dbar^0}\xspace}
\def\DzDzb   {\ensuremath{\Dz {\kern -0.16em \Dzb}}\xspace}
\def\Dp      {\ensuremath{\D^+}\xspace}
\def\Dm      {\ensuremath{\D^-}\xspace}

\def\DpDm    {\ensuremath{\Dp {\kern -0.16em \Dm}}\xspace}

\def\Bbar    {\ensuremath{\kern 0.18em\overline{\kern -0.18em \PB}{}}\xspace}

\def\jpsi     {\ensuremath{{\PJ\mskip -3mu/\mskip -2mu\Ppsi\mskip 2mu}}\xspace}

  \def\Y#1S{\ensuremath{\PUpsilon{(#1S)}}\xspace}% no space before {...}!

\def\Lbar {\ensuremath{\kern 0.1em\overline{\kern -0.1em\PLambda}}\xspace}

\def\to                 {\ensuremath{\rightarrow}\xspace}

\def\AT#1     {\ensuremath{A_{\mathrm{T}}^{#1}}\xspace}           % 2

\def\C#1      {\ensuremath{\mathcal{C}_{#1}}\xspace}                       % 9
\def\Cp#1     {\ensuremath{\mathcal{C}_{#1}^{'}}\xspace}                    % 7
\def\Ceff#1   {\ensuremath{\mathcal{C}_{#1}^{\mathrm{(eff)}}}\xspace}        % 9  
\def\Cpeff#1  {\ensuremath{\mathcal{C}_{#1}^{'\mathrm{(eff)}}}\xspace}       % 7
\def\Ope#1    {\ensuremath{\mathcal{O}_{#1}}\xspace}                       % 2
\def\Opep#1   {\ensuremath{\mathcal{O}_{#1}^{'}}\xspace}                    % 7

\newcommand{\tev}{\ensuremath{\mathrm{\,Te\kern -0.1em V}}\xspace}
\newcommand{\gev}{\ensuremath{\mathrm{\,Ge\kern -0.1em V}}\xspace}
\newcommand{\mev}{\ensuremath{\mathrm{\,Me\kern -0.1em V}}\xspace}
\newcommand{\kev}{\ensuremath{\mathrm{\,ke\kern -0.1em V}}\xspace}
\newcommand{\ev}{\ensuremath{\mathrm{\,e\kern -0.1em V}}\xspace}
\newcommand{\gevc}{\ensuremath{{\mathrm{\,Ge\kern -0.1em V\!/}c}}\xspace}
\newcommand{\mevc}{\ensuremath{{\mathrm{\,Me\kern -0.1em V\!/}c}}\xspace}
\newcommand{\gevcc}{\ensuremath{{\mathrm{\,Ge\kern -0.1em V\!/}c^2}}\xspace}
\newcommand{\gevgevcccc}{\ensuremath{{\mathrm{\,Ge\kern -0.1em V^2\!/}c^4}}\xspace}
\newcommand{\mevcc}{\ensuremath{{\mathrm{\,Me\kern -0.1em V\!/}c^2}}\xspace}

\def\mum  {\ensuremath{\,\upmu\rm m}\xspace}

\def\invfb   {\ensuremath{\mbox{\,fb}^{-1}}\xspace}

\def\gsim{{~\raise.15em\hbox{$>$}\kern-.85em
          \lower.35em\hbox{$\sim$}~}\xspace}
\def\lsim{{~\raise.15em\hbox{$<$}\kern-.85em
          \lower.35em\hbox{$\sim$}~}\xspace}

\def\sPlot{\mbox{\em sPlot}}

\def\pt         {\mbox{$p_{\rm T}$}\xspace}

\def\evtgen     {\mbox{\textsc{EvtGen}}\xspace}
\def\pythia     {\mbox{\textsc{Pythia}}\xspace}

\def\geant      {\mbox{\textsc{Geant4}}\xspace}

\def\gauss      {\mbox{\textsc{Gauss}}\xspace}

\def\photos     {\mbox{\textsc{Photos}}\xspace}

\def\tell1  {TELL1\xspace}
\def\ukl1   {UKL1\xspace}

\usepackage{cite} % Allows for ranges in citations
\usepackage{mciteplus}

\usepackage{longtable} % only for template; not usually to be used in PAPERs

\begin{document}

%%%%%%%%%%%%%%%%%%%%%%%%%
%%%%% Title     %%%%%%%%%
%%%%%%%%%%%%%%%%%%%%%%%%%
\renewcommand{\thefootnote}{\fnsymbol{footnote}}
\setcounter{footnote}{1}

% %%%%%%% CHOOSE TITLE PAGE--------
%\onecolumn
% \input{title-LHCb-ANA}
%\input{title-LHCb-CONF}
% $Id: title-LHCb-PAPER.tex 34385 2013-04-25 09:15:20Z needham $
% ===============================================================================
% Purpose: LHCb-PAPER journal paper title page template
% Author: 
% Created on: 2010-09-25
% ===============================================================================

%%%%%%%%%%%%%%%%%%%%%%%%%
%%%%%  TITLE PAGE  %%%%%%
%%%%%%%%%%%%%%%%%%%%%%%%%
\begin{titlepage}
\pagenumbering{roman}

% Header ---------------------------------------------------
\vspace*{-1.5cm}
\centerline{\large EUROPEAN ORGANIZATION FOR NUCLEAR RESEARCH (CERN)}
\vspace*{1.5cm}
\hspace*{-0.5cm}
\begin{tabular*}{\linewidth}{lc@{\extracolsep{\fill}}r}
\ifthenelse{\boolean{pdflatex}}% Logo format choice
{\vspace*{-2.7cm}\mbox{\!\!\!\includegraphics[width=.14\textwidth]{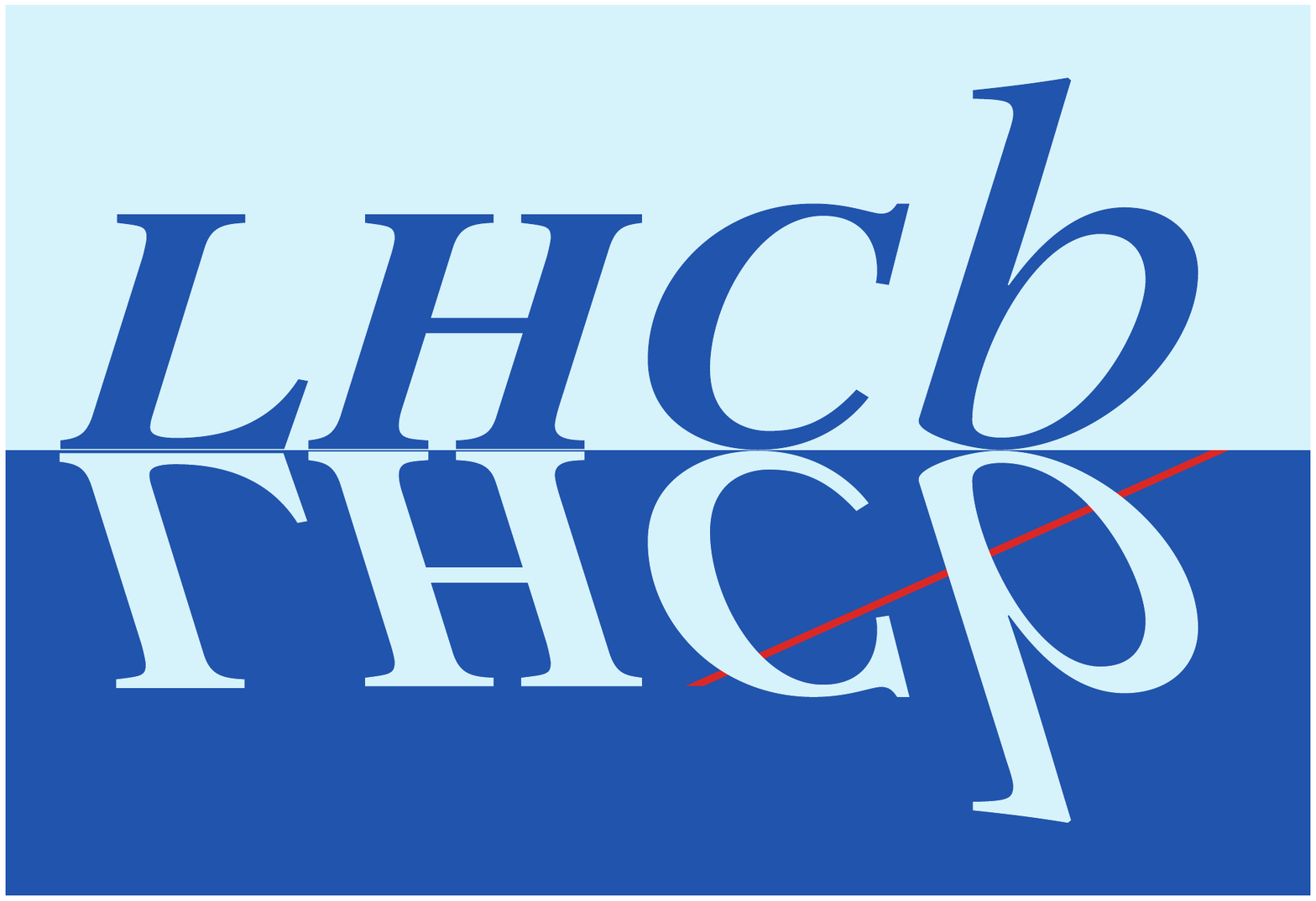}} & &}%
{\vspace*{-1.2cm}\mbox{\!\!\!\includegraphics[width=.12\textwidth]{lhcb-logo.eps}} & &}%
\\
 & & CERN-PH-EP-2013-053 \\  % ID 
 & & LHCb-PAPER-2013-011 \\  % ID 
 & & 27 May 2013
\end{tabular*}

\vspace*{4.0cm}

% Title --------------------------------------------------
{\bf\boldmath\huge
\begin{center}
  Precision measurement of $D$ meson mass differences
\end{center}
}

\vspace*{1.5cm}

% Authors -------------------------------------------------
\begin{center}
The LHCb collaboration\footnote{Authors are listed on the following pages.}
\end{center}
\vspace{\fill}
% Abstract -----------------------------------------------
\begin{abstract}
  \noindent
Using three- and four-body decays of $D$ mesons produced in semileptonic
$b$-hadron decays, precision measurements of $D$ meson mass differences are
made together with a measurement of the $D^{0}$ mass.  The
measurements are based on a dataset corresponding to an integrated
luminosity of $1.0~\invfb$ collected in $pp$ collisions at 7\,$\tev$. Using the decay
$D^0 \rightarrow K^{+} K^{-} K^{-} \pi^{+}$, the $D^0$ mass is measured to be
\begin{alignat*}{3}
M(D^0) \phantom{ghd} &=&~1864.75 \pm 0.15 \,({\rm stat}) \pm 0.11
\,({\rm syst}) \, \mevcc.
\end{alignat*}
The mass differences
\begin{alignat*}{3}
M(D^{+}) - M(D^{0})  &=& 4.76 \pm 0.12 \,({\rm stat}) \pm 0.07
\,({\rm syst}) \, \mevcc , \\
M(D^{+}_s) - M(D^{+}) &=& \phantom{00}98.68 \pm 0.03 \,({\rm stat}) \pm
0.04  \,({\rm syst}) \, \mevcc
\end{alignat*}
are measured using the $D^0 \rightarrow K^{+} K^{-} \pi^{+} \pi^{-}$
and $D^{+}_{(s)} \rightarrow K^{+}K^{-} \pi^{+}$ modes.
\end{abstract}

\vspace*{1.0cm}

\begin{center}
  Submitted to JHEP
\end{center}

\vspace{\fill}

{\footnotesize 
\centerline{\copyright~CERN on behalf of the \lhcb collaboration, license \href{http://creativecommons.org/licenses/by/3.0/}{CC-BY-3.0}.}}
\vspace*{2mm}

\end{titlepage}

%%%%%%%%%%%%%%%%%%%%%%%%%%%%%%%%
%%%%%  EOD OF TITLE PAGE  %%%%%%
%%%%%%%%%%%%%%%%%%%%%%%%%%%%%%%%

%  empty page follows the title page ----
\newpage
\setcounter{page}{2}
\mbox{~}
\newpage

% Author List ----------------------------
%  You need to get a new author list!
%%%%%%%%%%%%%%%%%%%%%%%%%%%%%%%%%%%%%%%%%%
\centerline{\large\bf LHCb collaboration}
\begin{flushleft}
\small
R.~Aaij$^{40}$, 
C.~Abellan~Beteta$^{35,n}$, 
B.~Adeva$^{36}$, 
M.~Adinolfi$^{45}$, 
C.~Adrover$^{6}$, 
A.~Affolder$^{51}$, 
Z.~Ajaltouni$^{5}$, 
J.~Albrecht$^{9}$, 
F.~Alessio$^{37}$, 
M.~Alexander$^{50}$
% ...

{\footnotesize \it
% ...
$ ^{5}$Clermont Universit\'{e}, Universit\'{e} Blaise Pascal, CNRS/IN2P3, LPC, Clermont-Ferrand, France\\
$ ^{6}$CPPM, Aix-Marseille Universit\'{e}, CNRS/IN2P3, Marseille, France\\
% ...
$ ^{9}$Fakult\"{a}t Physik, Technische Universit\"{a}t Dortmund, Dortmund, Germany\\
% ...
$ ^{35}$Universitat de Barcelona, Barcelona, Spain\\
$ ^{36}$Universidad de Santiago de Compostela, Santiago de Compostela, Spain\\
$ ^{37}$European Organization for Nuclear Research (CERN), Geneva, Switzerland\\
% ...
$ ^{40}$Nikhef National Institute for Subatomic Physics, Amsterdam, The Netherlands\\
% ...
$ ^{45}$H.H. Wills Physics Laboratory, University of Bristol, Bristol, United Kingdom\\
% ...
$ ^{50}$School of Physics and Astronomy, University of Glasgow, Glasgow, United Kingdom\\
$ ^{51}$Oliver Lodge Laboratory, University of Liverpool, Liverpool, United Kingdom\\
% ...
$ ^{n}$LIFAELS, La Salle, Universitat Ramon Llull, Barcelona, Spain\\
% ...
}
\end{flushleft}
%%%%%%%%%%%%%%%%%%%%%%%%%%%%%%%%%%%%%%%%%%

\cleardoublepage

%\twocolumn
% %%%%%%%%%%%%% ---------

\renewcommand{\thefootnote}{\arabic{footnote}}
\setcounter{footnote}{0}

%%%%%%%%%%%%%%%%%%%%%%%%%%%%%%%%
%%%%%  Table of Content   %%%%%%
%%%%%%%%%%%%%%%%%%%%%%%%%%%%%%%%
%%%% Uncomment next 2 lines if desired
%\tableofcontents
%\cleardoublepage

%%%%%%%%%%%%%%%%%%%%%%%%%
%%%%% Main text %%%%%%%%%
%%%%%%%%%%%%%%%%%%%%%%%%%

\pagestyle{plain} % restore page numbers for the main text
\setcounter{page}{1}
\pagenumbering{arabic}

%% Uncomment during review phase. 
%% Comment before a final submission.
%\linenumbers

% You can include short sections directly in the main tex file.
% However, for larger papers it is desirable to split the text into
% several semiautonomous files, which can be revised independently.
% This is especially useful when developing a document in
% collaboration with several people, since then different parts can be
% edited independently.  This type of file organization is shown here.
% 

\section{Introduction}
\label{sec:introduction}
Mesons are colourless objects composed of a quark-antiquark pair 
bound via the strong interaction. Measurements of meson masses provide
observables that can be compared to theoretical predictions. For 
the case of $B$ mesons, precision measurements
have been reported in recent years by several experiments~\cite{babar, CDFMasses, LHCb-PAPER-2011-035}. In contrast,
few precision $D$ meson mass measurements exist.

 For the $D^0$ meson\footnote{The inclusion of charge conjugate states
 is implied.}  the current average of $M(D^0) = 1864.91 \pm 0.17 \,\mevcc$, quoted by
the Review of Particle Physics \cite{PDG2012}, is dominated by the
measurements of the CLEO
\cite{Cawlfield:2007dw} and KEDR 
\cite{Anashin:2009aa} collaborations. Current knowledge of the masses  
of the $D^{+}$ and $D^{+}_s$  mesons, and the mass splitting between 
these states, is more limited.  The most precise 
determination of the  $D^{+}$ mass is made by the KEDR collaboration
\cite{Anashin:2009aa}  resulting in $M(D^+) = 1869.53 \pm
0.49\,\mathrm{(stat)}\, \pm 0.20\,\mathrm{(syst)} \,\mevcc$. In
addition, two measurements of the mass
splitting between the $D^{+}$ and $D^{0}$ mesons by the MRK2
\cite{mrk2} and LGW \cite{lgw} collaborations have been reported. These are
averaged \cite{PDG2012} to give \mbox{$M(D^{+}) - M(D^{0}) = 4.76 \pm 0.28 \,\mevcc$}. No absolute measurement of the $D^{+}_s$ mass with a precision
better than the $\mevcc$ level exists and the reported values 
are not in good agreement \cite{PDG2012}. More precise measurements of
the mass difference relative to the $D^{+}$ meson have been reported by
several collaborations \cite{Acosta:2003qr,Aubert:2002ue,Brown:1994gu, Chen:1989tu, Anjos:1987eq}. These are averaged \cite{PDG2012} to give $M(D^{+}_s)$ $-$ $M(D^{+}) = 98.85 \pm
0.25 \,\mevcc$. The fit of open charm mass data \cite{PDG2012} leads 
to $M(D^{+}_s) =  1968.49 \pm 0.32\,\mevcc$. Though this value is
significantly more precise than the direct measurement, it would still
dominate the systematic uncertainty on the measurement of the
$B_c^{+}$ mass in the $B_c^+ \rightarrow  J/\psi D^{+}_s$ decay mode \cite{LHCb-PAPER-2013-010}. 

Recent interest in the $D^0$ mass has been driven by the 
observation of the $X(3872)$ state, first measured by the Belle experiment 
\cite{Choi:2003ue} and subsequently confirmed elsewhere \cite{CDFPhysRevLett.93.072001, 
D0Abazov:2004kp,
BaBarPhysRevD.71.071103,LHCb-PAPER-2011-034,Chatrchyan:2013cld} . This
state, with $J^{PC} = 1^{++}$ \cite{LHCb-PAPER-2013-001}, does not fit well into the quark model picture, and exotic 
interpretations have been suggested: for example that it is a
tetraquark \cite{Maiani:2004vq} or a loosely bound deuteron-like
 $D^{*0}{\overline{D}}{}^0$ `molecule' \cite{Tornqvist:2004qy}. For the
 latter interpretation to be 
valid, the mass of the $X(3872)$ state should be less than the sum of 
the $D^{*0}$ and ${D}^0$ masses. Using the fitted value of the $D^0$
mass and the measured values for the other quantities quoted in Ref.~\cite{PDG2012}, the binding energy
($E_B$) in this interpretation can be estimated to be
\begin{eqnarray}
E_\textrm{B} &=& M(D^{0}D^{*0}) - M(X(3872))   \nonumber \\
       &=& 2M(D^{0}) + \Delta M(D^{*0}-{D}^0) - M(X(3872))   \nonumber \\
      &=& 0.16 \pm 0.32 \mevcc.  \nonumber
\end{eqnarray}
Therefore, the issue of whether the $X(3872)$ can be a bound 
molecular state remains open. To clarify the 
situation, more precise measurements of both the $X(3872)$ and $D^0$ masses are needed.

In this paper, a measurement of the $D^{0}$ mass using the 
$D^0 \rightarrow K^{+} K^{-} K^{-} \pi^{+}$ decay mode is
reported. This mode has a relatively low energy release, $Q$-value, 
defined as the difference between the mass of the $D$ meson and the sum of
the masses of the daughter particles. Consequently, systematic uncertainties due 
to the calibration of the momentum scale of the detector are reduced. Other four-body
$D^{0}$ decay modes are used to provide a cross-check of the
result. In addition, precision measurements of the $D^{+}-{D}^0$ and $D^{+}_s-{D}^+$ mass
differences are made. For the mass difference measurements the 
$D^0 \rightarrow K^{+} K^{-} \pi^{+} \pi^{-}$ mode is used, together
with the $D^{+}_{(s)} \rightarrow K^{+}K^{-} \pi^{+}$ decay, since
these modes have similar $Q$-values. 

\section{Detector and dataset}
\label{sec:dataset}
The analysis uses data, corresponding to an integrated luminosity of
$\rm 1.0~fb^{-1}$, collected in $pp$ collisions at a centre-of-mass
energy of $\sqrt{s}$\,=\,7\,$\tev$ by the \lhcb experiment during
2011. The detector response is studied using
a simulation. Proton-proton collisions are generated using
\pythia~6.4~\cite{Sjostrand:2006za} with the configuration described
in Ref~\cite{LHCb-PROC-2010-056}. 
Particle decays are then simulated by \evtgen~\cite{Lange:2001uf} 
in which final state radiation is generated using
\photos~\cite{Golonka:2005pn}. The 
interaction of the generated particles with the detector and its 
response are implemented using the \geant 
toolkit~\cite{Allison:2006ve, *Agostinelli:2002hh} with the settings
described in Ref.~\cite{LHCb-PROC-2011-006}.

The \lhcb detector~\cite{Alves:2008zz} is a single-arm forward
spectrometer covering the pseudo-rapidity range $2<\eta <5$. 
It includes a high precision tracking system consisting of a
silicon-strip vertex detector surrounding the $pp$ interaction region,
a large-area silicon-strip detector located upstream of a dipole
magnet with a bending power of about $4{\rm\,Tm}$, and three stations
of silicon-strip detectors and straw drift tubes placed
downstream. The polarity of the dipole magnet is reversed at intervals
that correspond to roughly $0.1 \, \invfb$ of collected data 
in order to minimize systematic uncertainties. 
The combined tracking system has momentum resolution
$\Delta p/p$ that varies from 0.4\% at 5\gevc to 0.6\% at 100\gevc,
and impact parameter resolution of 20\mum for tracks with high
transverse momentum ($\pt$). Charged hadrons are identified using two
ring-imaging Cherenkov detectors. Photon, electron and hadron
candidates are identified by a calorimeter system consisting of
scintillating-pad and pre-shower detectors, an electromagnetic
calorimeter and a hadronic calorimeter. Muons are identified by a
system composed of alternating layers of iron and multiwire
proportional chambers. The trigger \cite{LHCb-DP-2012-004} consists of a hardware stage, based
on information from the calorimeter and muon systems, followed by a
software stage that applies a full event reconstruction. Samples of
 open charm mesons produced directly in the primary $pp$ interaction (refered
to as `prompt') and in semileptonic decays of
$b$-hadrons are selected by the trigger. Though the prompt sample is
larger in size, cuts on the decay time of the $D$ meson are
applied at the trigger level to reduce the output rate. As the
reconstructed mass and decay time are correlated, these cuts bias the
mass measurement. In contrast, no cuts on the $D$ decay time are
applied at the trigger level for the semileptonic sample, which is
therefore used for this analysis.

The measurements require the momenta of the final state particles to
be determined accurately. The procedure used to calibrate the momentum
scale of the tracking system for this study is discussed in detail in 
Ref.~\cite{LHCb-PAPER-2012-048}. It is based upon large calibration samples of 
$B^+ \to \jpsi K^+$ and $\jpsi\to\mup\mun$ decays collected
concurrently with the dataset used for this analysis. The use of the
large $J/\psi$ dataset allows to correct for variations of the
momentum scale at the level of $10^{-4}$ or less that occur over the course of
the data-taking period whilst the use of the $B^+ \to \jpsi K^+$ 
allows the momentum scale to be determined as a function of the $K^+$ 
kinematics. The accuracy of the procedure has been checked using other
fully reconstructed $B$ decays together with two-body $\Upsilon(nS)$ and $\KS$
decays. In each case the deviation of the measured mass from the
expected value is converted to an estimate of the bias on the momentum
scale ($\alpha$) taking into account relativistic kinematics 
and QED radiative corrections. The
largest value of $|\alpha|$ found in these studies
is $0.03\,\%$ for the $\KS \rightarrow \pi^+ \pi^-$ decay 
mode. Conservatively, this is taken as the  uncertainty on the 
calibrated momentum scale. This leads to the largest contribution to
the systematic uncertainty on the mass measurements. 

\section{Selection}
\label{sec:selection}
The selection uses only well reconstructed charged
particles that traverse the entire tracking system.  All charged
particles are required to be within the angular acceptance
of the spectrometer. This corresponds to 300~mrad in the bending plane of
the dipole magnet and 250~mrad in the orthogonal plane. In addition,
the final state particles are required to have $\pt$ greater than
300~\mevc. Further background suppression is achieved by exploiting
the fact that the products of heavy flavour decays have a large
distance of closest approach (`impact parameter') with respect to the $pp$ interaction vertex in which they were
produced. The impact parameter $\chi^2$ with respect to any
primary vertex is required to be larger than nine. Fake tracks created by the
reconstruction are suppressed by cutting on the output of a neural network trained to
discriminate between these and real particles. This cut also removes candidates where one of
the charged hadrons has decayed in flight. To select well-identified
kaons (pions) the difference in the logarithms of the
global likelihood of the kaon (pion) hypothesis relative to the pion
(kaon) hypothesis provided by the ring-imaging Cherenkov detectors is required to be greater than five (zero).  

Charged particles selected in this way are combined to form $\D^{0}
\rightarrow K^{+} K^{-} \pi^{+} \pi^{-} $,  $\D^{0} \rightarrow K^{+}
K^{-} K^{-} \pi^{+} $ and  $\D^{+}_{(s)} \rightarrow K^{+} K^{-}
\pi^{+} $  candidates. To eliminate kinematic reflections due to 
misidentified pions, the invariant mass of at least one kaon pair is required to be within $ \pm 12~\mevcc$ of the 
nominal value of the $\Pphi$~meson mass \cite{PDG2012}. This requirement
means that the $D$ meson sample is dominated by decays containing an
intermediate~$\Pphi$ meson. A fit requiring the final state particles to originate from a common point is made and the 
$\chi^2$ per degree of freedom ($\chi^2/$ndf) of this fit is required to be less than 
five. In order to remove poorly reconstructed candidates, a cut is made
on the uncertainty of the reconstructed invariant mass estimated by 
propagation of the individual track covariance matrices. The value of 
this cut depends on the decay mode under consideration and is chosen such that the bulk of the
 distribution is kept and only events in the tail are
 rejected. In a few percent of the events the reconstruction procedure gives rise to
 duplicate candidates. Therefore, if two or more candidates 
that are separated by less than 0.05 in pseudorapidity and $50~$mrad in azimuthal
angle are found within one event, only that with the best $D$ vertex $\chi^2$ is kept. 

Each candidate $D$ meson, selected in this way, is combined with a
well-identified muon that is displaced from the $pp$ interaction vertex (impact parameter $\chi^2 > 4$) and that has $\pt$ larger than $800~\mevc$ to form a $B$ candidate. A fit is made 
requiring the muon and the $D$ candidate to originate from a common point and 
the $\chi^2$ per degree of freedom of this fit is required to be less
than five.  To select semileptonic $B$ decays, the invariant
mass of the $B$ candidate is required to be in the range $2.5 -
6.0~\gevcc$. In principle, the large combinatorial background can be further reduced by cutting on the decay time of the $D$~meson, but due to the correlation between the decay time and the
mass, this cut would bias the mass distribution. Therefore, a cut requiring significant
displacement between the $b$-hadron decay vertex and the associated $pp$ interaction
vertex is applied. This achieves high signal purity whilst not biasing
the distribution of the $D$ decay time.

\section{Fit Results}      
\label{sec:fitresults}   
The $D$ meson masses are determined by performing extended unbinned maximum
likelihood fits to the invariant mass distributions. In these fits the
background is modelled by an exponential function and the signal by
the sum of a Crystal Ball \cite{Skwarnicki:1986xj} and a Gaussian function. The Crystal Ball
component accounts for the presence of the QED radiative tail. Alternative models
for both the signal and background components are considered as part
of the studies of the systematic uncertainties. The model for the signal shape contains six parameters: 
\begin{itemize}
\item a common mean value for the Gaussian and Crystal Ball components;
\item the widths of the Gaussian  ($\sigma_{\mathrm{G}}$) and the  Crystal Ball
  ($\sigma_{\mathrm{CB}}$) components;
\item the transition point ($a$) and exponent ($n$) of the Crystal
  Ball component;
\item the relative fraction of the Crystal Ball ($f_{\textrm{CB}}$) component.
\end{itemize}
To reduce the number of free parameters in the fit, $a$, $n$ and $f_{\mathrm{CB}}$
together with the ratio of $\sigma_{\mathrm{CB}}$ to $\sigma_{\mathrm{G}}$, are fixed using a simulation that
has been tuned to reproduce the mass resolution observed in data for the $B^+ \to
J/\psi  K^+$ and $B^+ \to J/\psi  K^+ \pi^{-} \pi^{+}$
decay modes. By fixing the ratio of $\sigma_{\mathrm{CB}}$ to
$\sigma_{\mathrm{G}}$ the resolution model is constrained up to an overall
resolution scale factor that is close to unity. 

The Crystal Ball function describes the effect of the radiative tail far
from the peak well. However, close to the peak its shape is still
Gaussian, which results in a bias on the fitted mass that scales with the
$Q$-value of the decay mode. This effect is studied using \photos \cite{Golonka:2005pn} to model the effect of QED radiative
corrections. The size of the bias is found to be $0.03 \pm 0.01\,\mevcc$ for the $\D^{0} \rightarrow K^{+} K^{-} K^{-} \pi^{+} $
mode. For the $\D^{0} \rightarrow K^{+} K^{-} \pi^{+} \pi^{-} $, $\D^{+} \rightarrow K^{+} K^{-} \pi^{+} $ 
and $\D_{s}^{+} \rightarrow K^{+} K^{-} \pi^{+} $ decay modes a value of  $0.06
\pm 0.01~\mevcc$ is found. These values are used to correct the mass
measurements. The effect cancels in the measurement of the mass differences.

\begin{figure}[t!]
\centering
\includegraphics[scale=0.38]{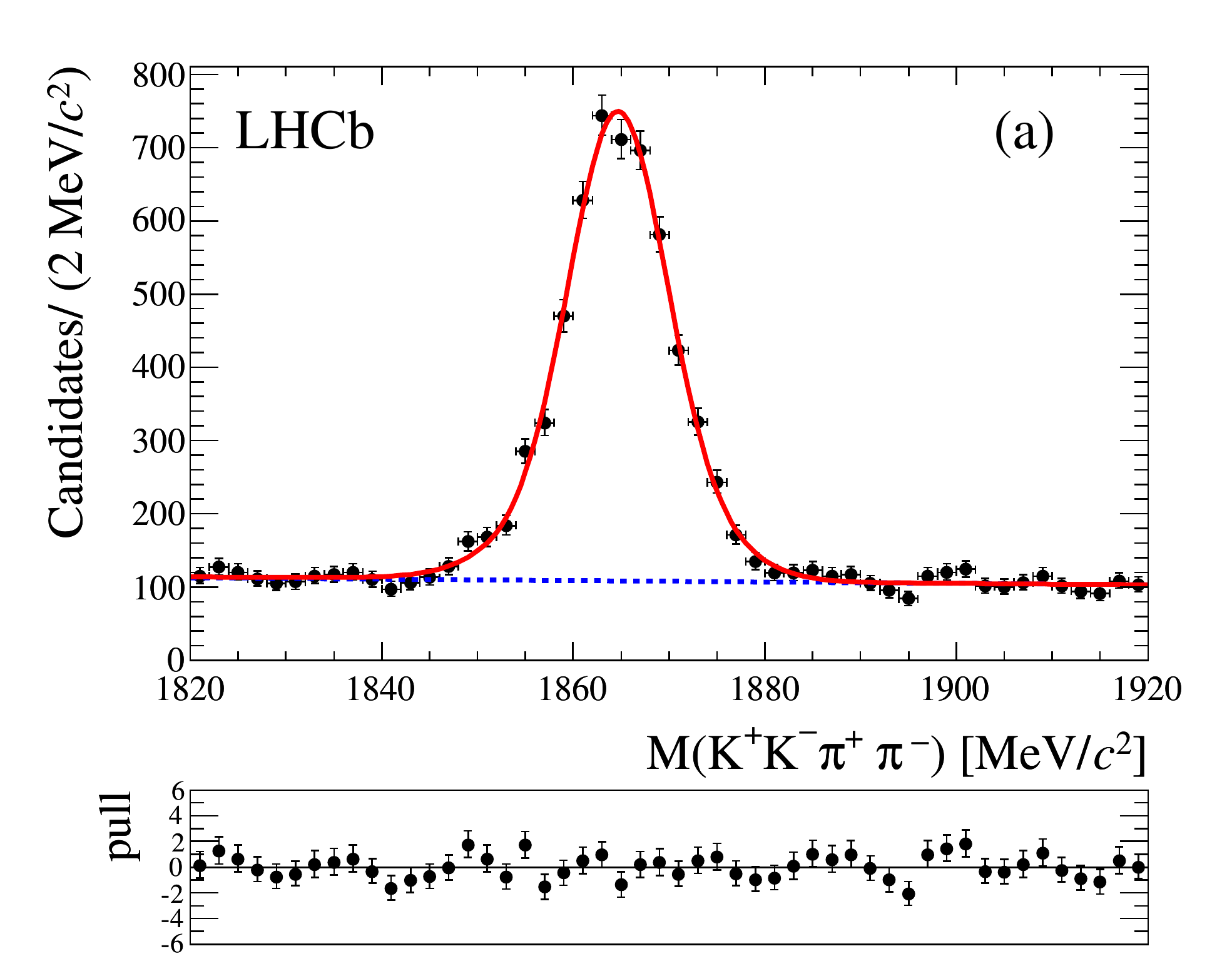}
\includegraphics[scale=0.38]{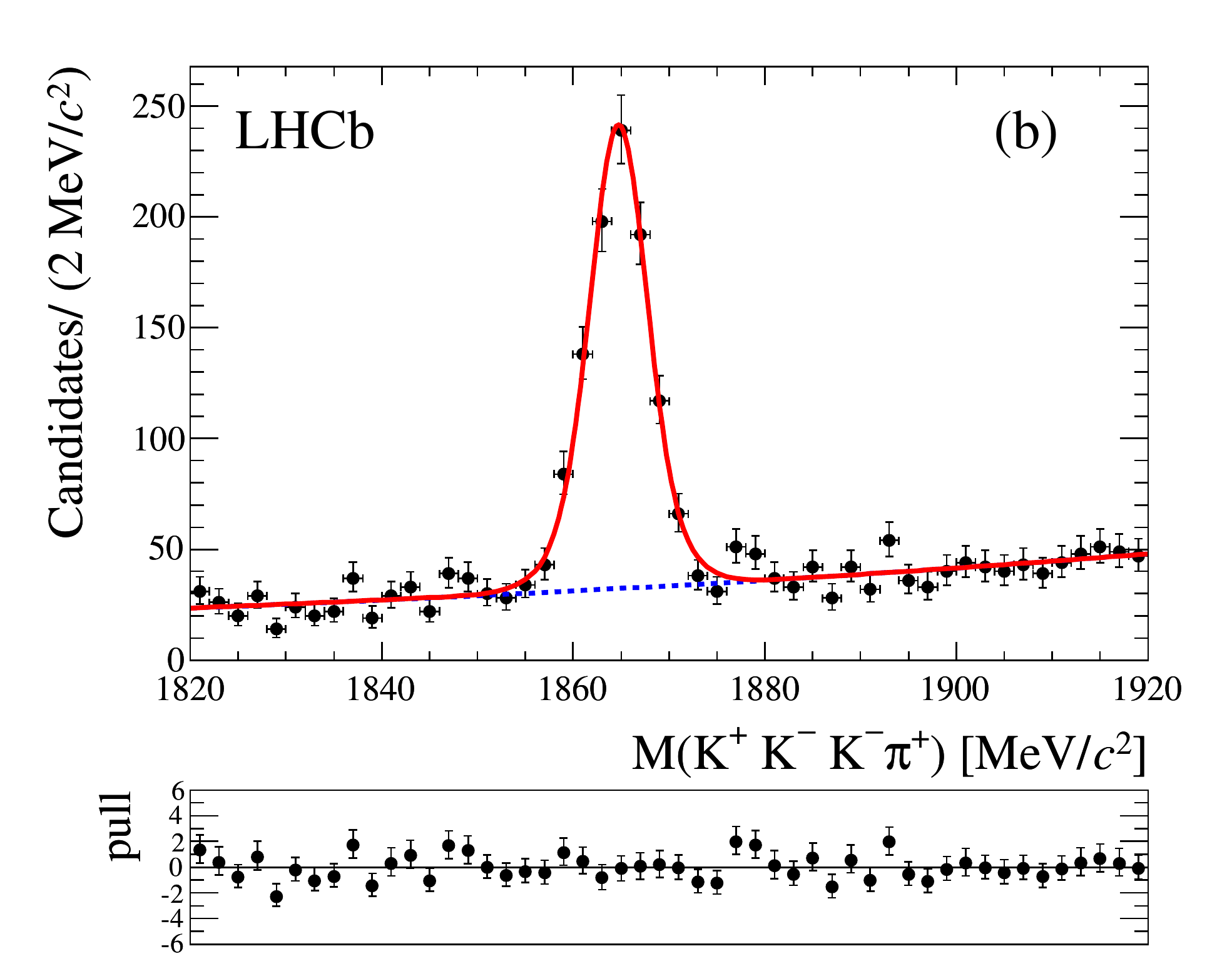} 
\caption{\small Invariant mass distributions for  
the (a) $K^{+} K^{-} \pi^{+} \pi^{-} $  
and (b) $K^{+} K^{-} K^{-} \pi^{+}$ final states. In each case
the result of the fit described in the text is
superimposed (solid line) together with the background component
(dotted line). The pull, i.e. the difference
between the fitted and measured value divided by the
uncertainty on the measured value, is shown below each plot.}
\label{fig:d0Masses}
\end{figure}

\begin{figure}[htb!]
\begin{center}
\resizebox{4.5in}{!}{\includegraphics{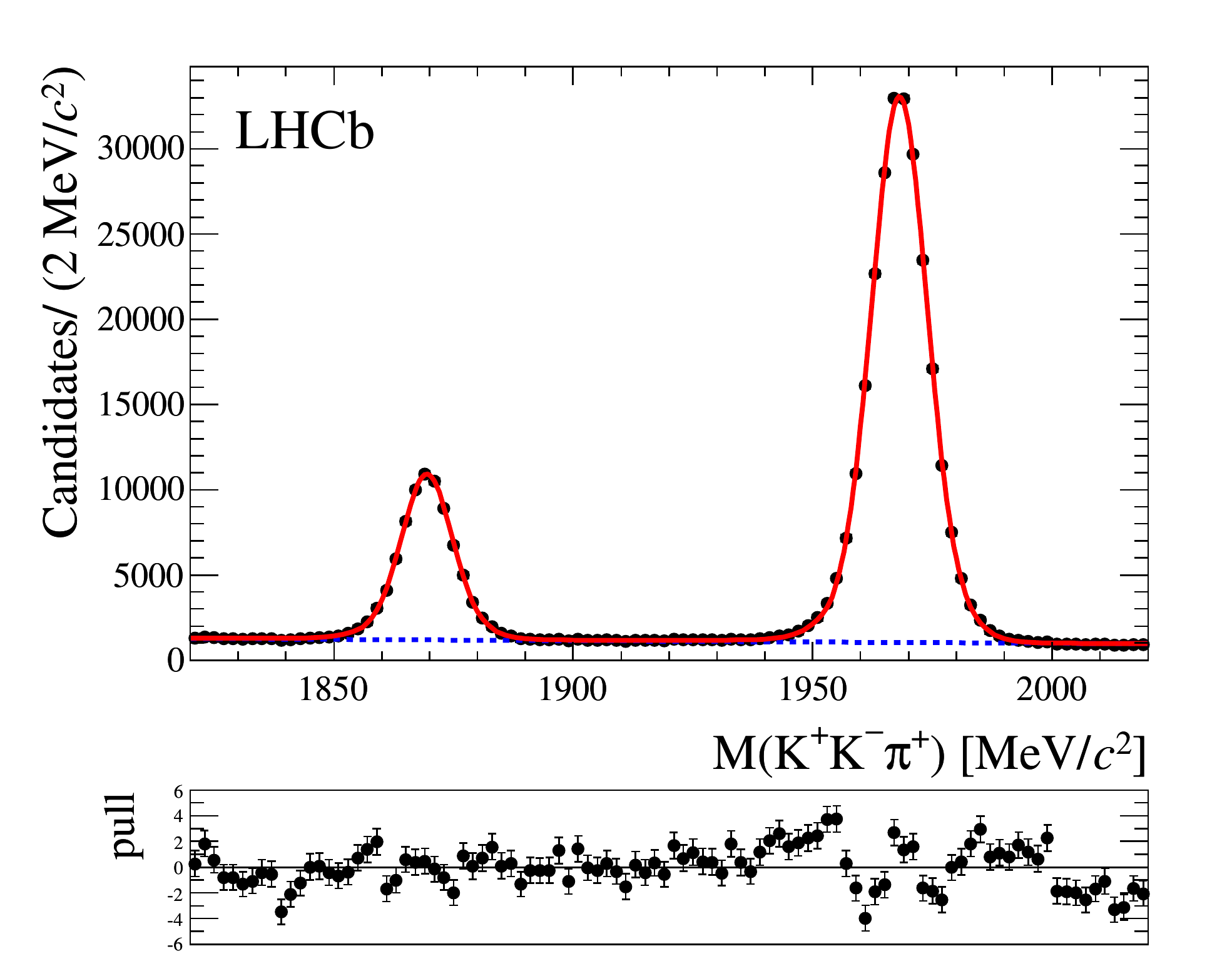}}
\vspace{-5mm}
\caption{\small Invariant mass distribution for  
the $K^{+} K^{-} \pi^{+}$ final state. The
result of the fit described in the text is 
superimposed (solid line) together with the background component 
(dotted line). The pull, i.e. the difference
between the fitted value and the measured value divided by the 
uncertainty,  is shown below the plot.}
\label{fig:dplusMasses}
\end{center}
\end{figure}

The resulting fits for the $D^0$ decay modes
are shown in Fig.~\ref{fig:d0Masses} and that for the  $K^{+} K^{-}
\pi^+$ final state in Fig.~\ref{fig:dplusMasses}. The values obtained
in these fits are summarized in  Table~\ref{tab:yields}. The resulting
values of the $D^+$ and $D^+_s$ masses 
are in agreement with the current world averages. These modes
have relatively large $Q$-values and consequently the 
systematic uncertainty due to the knowledge of the momentum scale 
is at the level of $0.3\,\mevcc$. Hence, it is chosen not to
quote these values as measurements. Similarly, the systematic
uncertainty due to the momentum
scale for the $\D^{0} \rightarrow K^{+} K^{-} \pi^{+} \pi^{-}$
mode is estimated to be $0.2~\mevcc$ and the measured mass
in this mode is not used in the $\D^0$ mass determination. 
\begin{table}[t!]
\caption{\small Signal yields, mass values, resolution scale factors
  and binned $\chi^2$/ndf (using 100 bins) obtained from the fits shown in Fig.~\ref{fig:d0Masses} and Fig.~\ref{fig:dplusMasses} together with
  the values corrected for the effect of QED radiative corrections as
  described in the text.}
\begin{center}
%\small
\resizebox{\textwidth}{!}{
\begin{tabular}{l|c|c|c|c|c}
 &  &  Fitted mass  &  Corrected mass  & Resolution  & \\ 
 \raisebox{1.5ex}[-1.5ex]{Decay mode} &
 \raisebox{1.5ex}[-1.5ex]{Yield} &[$\!\mevcc$] & [$\!\mevcc$]  &
 scale factor  & \raisebox{1.5ex}[-1.5ex]{$\chi^2$/ndf} \\ \hline
$\D^{0} \, \rightarrow K^{+} K^{-} \pi^{+} \pi^{-} $      &
$\phantom{111}4608\pm89\phantom{1} $ & $1864.68 \pm 0.12$ & $1864.74 \pm 0.12$ & $1.031
\pm 0.021$ & 0.83 \\
$\D^{0} \, \rightarrow K^{+} K^{-} K^{-} \pi^{+} $        & $\phantom{1111}849
\pm36\phantom{1} $ & $1864.73 \pm 0.15$  & $1864.75 \pm 0.15$ & $0.981 \pm
0.042$ & 0.92 \\
$\D^{+} \rightarrow K^{+} K^{-}  \pi^{+} $        & $\phantom{1}68,787 \pm 321$ & $1869.44 \pm 0.03$  & $1869.50 \pm 0.03$ & $0.972 \pm
0.003$ &  \\
$\D^{+}_s \rightarrow K^{+} K^{-}  \pi^{+} $        & $248,694\pm 540$ & $1968.13 \pm 0.03$  & $1968.19 \pm 0.03$ & $0.971 \pm
0.002$ & \raisebox{1.5ex}[-1.5ex]{2.5} \\
\end{tabular}}
\end{center}
\label{tab:yields}
\end{table}

The quality of the fits is judged from the $\chi^2$/ndf, quoted in Table~\ref{tab:yields}, and the fit
residuals. It has been checked using simulated pseudo-experiments 
that the sizeable trends seen in the residuals for the $K^+ K^- \pi^+$
mode, where the dataset is largest, do not bias the 
mass difference measurement. The fitted resolution scale factors are all within a
few percent of unity, indicating that the calibration parameters obtained from
the $B^+$ study are applicable in this analysis. The uncertainties on
the masses reported by the fits are in good agreement with the results obtained in
pseudo-experiments.

Using the values in Table~\ref{tab:yields}, the mass differences are evaluated to be
\begin{center}
\begin{tabular}{l @{$~=~$}l@{$\,\,\pm\,$}l@{\,(stat) MeV/$c^2$}l}
$M(D^{+}) - M(D^{0})$   & \phantom{1}4.76 & 0.12 &,  \\
$M(D^{+}_s) - M(D^{+})$   & 98.68 & 0.03 & , \\
\end{tabular}
\end{center}
where the uncertainties are statistical only.

\section{Systematic uncertainties}
\label{sec:systematics}
To evaluate the systematic uncertainty, the complete analysis is
repeated, including the track fit and the
momentum scale calibration when needed, varying within their
uncertainties the parameters to which the mass determination is
sensitive. The observed changes in the central values of the 
fitted masses relative to the nominal results are assigned as systematic
uncertainties. 

The dominant source of uncertainty is  the
limited knowledge of the momentum scale. The mass fits are repeated with the momentum
scale varied by $\pm 0.03\,\%$. A further uncertainty is related to
the understanding of the energy loss in the material of the tracking system. The amount of material traversed in the tracking system by a particle is 
known to 10\,\% accuracy~\cite{LHCb-PAPER-2010-001}. Therefore, the magnitude of the energy 
loss correction in the reconstruction is varied by $\pm10\,\%$. 

Other uncertainties arise from the fit model. To evaluate the impact 
of the signal model, a fit is performed where all signal parameters are
fixed according to the values found in the simulation and a second fit
where the parameters $\sigma_{\mathrm{G}}$ and 
$\sigma_{\mathrm{CB}}$ are allowed to vary while keeping the relative
fraction, $f_{\mathrm{CB}}$, of the two components fixed. The larger of the differences to the
default fit result is assigned as an estimate of the systematic
uncertainty. Similarly the effect of the background modelling is
estimated by replacing the exponential function with a first-order Chebychev
polynomial. The shifts of the mass values observed in these tests are generally much
smaller than $0.01\,\mevcc$ and are assigned as systematic uncertainties. For the $K^+K^-\pi^+$ fit further 
cancellation occurs in the mass difference. It is concluded that the details of the fit
model have little effect on the presented measurements. 

An additional uncertainty arises from the knowledge of the
value of the $K^{+}$ mass, $m_{K^{\pm}} = 493.677 \pm 0.016 \mevcc$ \cite{PDG2012}. 
The effect of this uncertainty on the measurements has been evaluated from simulation studies. 

The systematic uncertainties on the
measured masses and mass differences
are summarized in Table~\ref{tab:MassSyst}. The uncertainties related
to the momentum scale and energy loss correction are fully correlated
between the measurements. 
\begin{table}[t]
\caption{\small Systematic uncertainties (in \!\mevcc) on the 
mass measurements and on their differences.}
\label{tab:MassSyst}
\begin{center}
\small
\begin{tabular}{l|c|c|c}
Source of uncertainty & $M(D^{0})$ & $M(D^{+}) - M(D^{0})$ & $M(D^{+}_s) - M(D^{+})$ \\
\hline
 Momentum scale                  & \phantom{$<$}0.09  & \phantom{$<$}0.04 & \phantom{$<$}0.04 \\
Energy loss correction              & \phantom{$<$}0.03  & \phantom{$<$}0.06 & $<$0.01  \\
$K^{\pm}$ mass                 & \phantom{$<$}0.05 & $<$0.01  & $<$0.01   \\
Signal model                            & \phantom{$<$}0.02  & $<$0.01 & $<$0.01  \\
Background model                        & $<$0.01 & $<$0.01 & $<$0.01  \\
\hline
Quadratic sum                           & \phantom{$<$}0.11 & \phantom{$<$}0.07 & \phantom{$<$}0.04  \\
\end{tabular}
\end{center}
\end{table}

Various cross-checks of the measurements are made. Two checks are 
related to the knowledge of the tracking system
alignment. First, a study has been performed where particle
trajectories are reconstructed without using the information
related to the tracking detector located before the entrance of the
spectrometer magnet. This information is not required to form a track but improves 
the momentum resolution by $10-20\,\%$. The second test is 
to vary the track slopes in the vertex detector by the uncertainty of $2
\times 10^{-4}$ on the length scale of the detector described in 
Ref.~\cite{LHCb-PAPER-2013-006}. The results obtained in these studies 
are consistent with those presented here and no additional uncertainty is
assigned.  

\begin{figure}[htb!]
\begin{center}
\resizebox{5.5in}{!}{\includegraphics{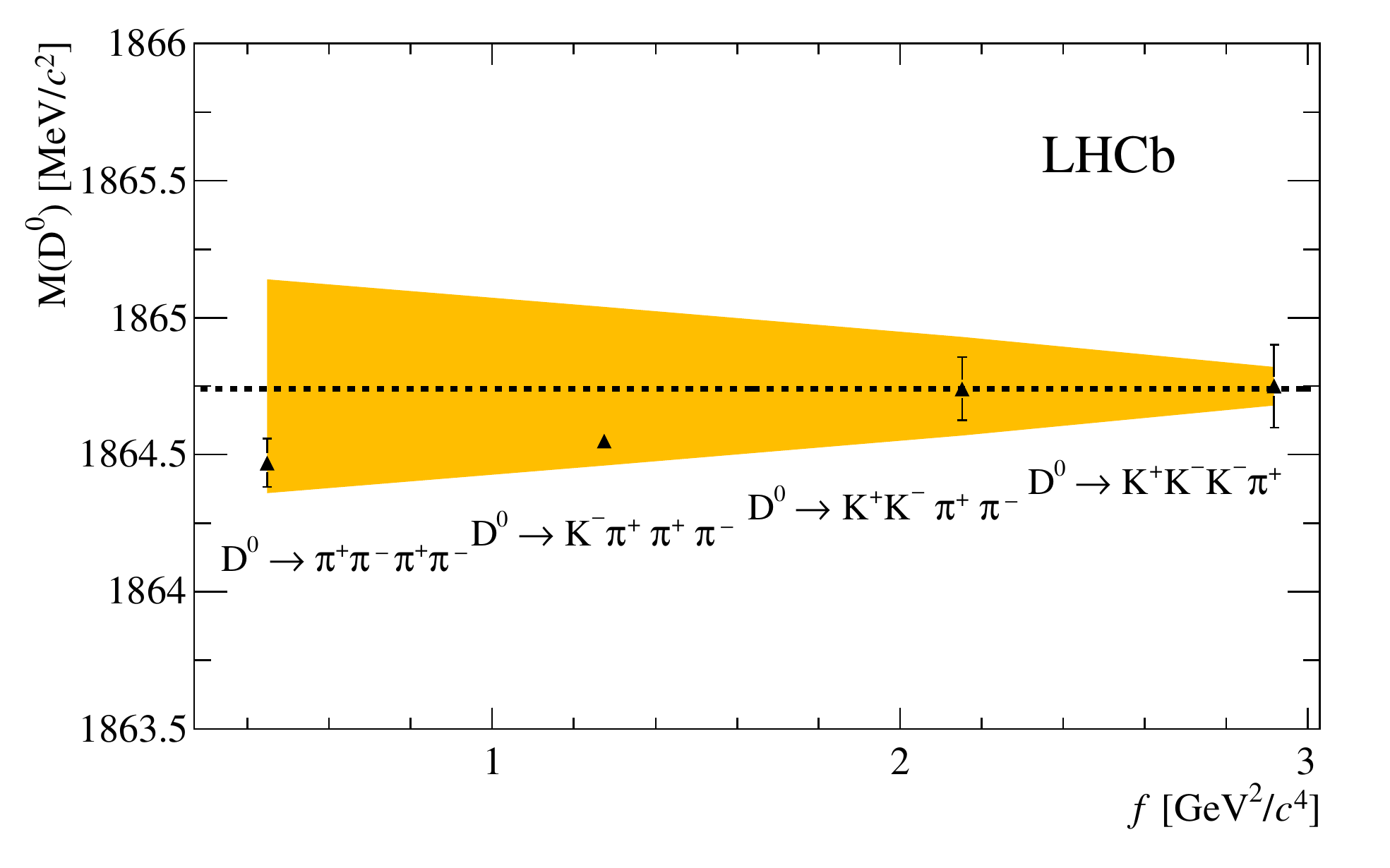}}
\vspace{-5mm}
\caption{\small Measured $D^0$ mass versus $f$ as defined in
  Eq.~\ref{fdef}. The (yellow) shaded area corresponds to a
  systematic uncertainty on the momentum scale of $0.03\,\%$ centred on
  the result for the $D^0 \rightarrow K^{+} K^{-} K^{-}
  \pi^{+}$ mode (horizontal dashed line). Only the $D^0 \rightarrow K^{+} K^{-} K^{-}
  \pi^{+}$ mode, where the systematic uncertainty is lowest, is used to
  determine the $D^0$ mass.}
\label{fig:fMass}
\end{center}
\end{figure}

A further check for the $D^0$ mass measurement is the comparison of the
measured mass in the $D^{0} \rightarrow K^+K^-K^-\pi^+ $ mode with that
obtained in the three other four-body modes. Systematic
effects related to the momentum scale will affect modes with a high $Q$-value more
than those with low $Q$-values. The relationship between the reconstructed
mass ($m$) and the momentum scale ($\alpha$) after a first-order
Taylor expansion in $m^2/p^2$ is
\begin{equation}
 m^2 =  \frac{m^2_{\mathrm{true}} - f }{(1-\alpha)^2}  + f,
\label{masseq}
\end{equation}
where 
\begin{equation}
f = p \sum \frac{  m_{i}^2 }  {p_{i}},
\label{fdef}
\end{equation}
$p$ is the total momentum of the decaying meson and $p_i$ and $m_i$  are the
momenta and masses of the daughter particles. This formalism assumes
that there are no additional differences affecting
the momentum scale between the modes such as differences in track kinematics or the effect of QED radiative corrections. For each decay
mode the average value of $f$ is obtained from the data using the
\sPlot\ technique \cite{Pivk:2004ty} with the mass as the control
variable to subtract the 
effect of background. The values obtained in this way are in good 
agreement with those found in the simulation.  In
Fig.~\ref{fig:fMass} the measured $D^{0}$ mass is plotted versus $f$
for the four-body decay modes studied here.  The shaded area on this
plot corresponds to the assigned systematic uncertainty of $0.03\,\%$
on the momentum scale. Though there is evidence of a systematic effect for the low
$f$-value modes it is accounted for by the assigned uncertainty.

The dataset has also been divided according to the magnet 
polarity and data-taking period and for the charged modes 
by the sign of the product of the magnet polarity and the $D$ meson 
charge. In addition, for modes where the event samples are sizable
the measurements are repeated in bins of the $D$ meson kinematic
variables. None of these tests reveal any evidence of a systematic bias.  

\section{Summary}
\label{sec:summary} 
Measurements of $D$ meson masses and mass differences have been
performed using $pp$ collision data, corresponding to an integrated
luminosity of $1.0~\invfb$ collected at a centre-of-mass energy of
$\sqrt{s} = 7 \, \tev$ with the LHCb detector. The results are
\begin{center}
\begin{tabular}{l @{$~=~$}l@{$\,\,\pm\,$}l@{\,(stat) $\pm\,$}l@{\,(syst) MeV/$c^2$}l}
$M(D^0)$  & 1864.75 & 0.15 & 0.11 & , \\
$M(D^{+})$ $-$ $M(D^{0})$   & \phantom{111}4.76 & 0.12 & 0.07 & , \\
$M(D^{+}_s)$ $-$ $M(D^{+})$   & \phantom{11}98.68 & 0.03 & 0.04 & . \\
\end{tabular}
\end{center}
The dominant systematic uncertainty is related to the knowledge of
the momentum scale.

As shown in Table~\ref{tab:summary}, these measurements
are in agreement with previous measurements. The results for
the mass differences have smaller uncertainty than any previously
reported value. The measured value of the $D^0$ mass has a similar precision to the
published CLEO result \cite{Cawlfield:2007dw}. Including this result in
the determination of the $X(3872)$ binding energy given in
Section~\ref{sec:introduction} gives $E_{B} = 0.09 \pm 0.28
\mevcc$. This reinforces the conclusion that if 
the $X(3872)$ state is a molecule it is extremely loosely bound.
 
\begin{table}[t]
\caption{\small LHCb measurements, compared to the best previous
  measurements and to the results of a global fit to available open charm mass data.
The quoted uncertainties are the quadratic sums of the statistical and systematic contributions. All values are in\,\mevcc.}
\vspace{2ex}
\begin{center}
\begin{tabular}{c|c|c|c}
 & LHCb  & Best previous &  \\ 
\raisebox{1.5ex}[-1.5ex]{Quantity} & measurement  & measurement   &
\raisebox{1.5ex}[-1.5ex]{PDG fit \cite{PDG2012}} \\
\hline
$M(D^0)$         & $1864.75 \pm 0.19$  & $1864.85 \pm 0.18$~~\cite{Cawlfield:2007dw}  & $1864.86 \pm 0.13$ \\
$M(D^+) -M(D^0)$  &  $\phantom{111}4.76\pm 0.14 $ &
$\phantom{111}4.7\phantom{0} \pm 0.3\phantom{0}$~~\cite{mrk2} & $~~~~~4.76 \pm 0.10$ \\   
$M(D^{+}_s) - M(D^{+})$  & $\phantom{11}98.68 \pm 0.05$ & $\phantom{111}98.4\phantom{0} \pm 0.3\phantom{0}$~~\cite{Aubert:2002ue} & $~~~98.88 \pm 0.25 $  \\  
\end{tabular}
\end{center}
\label{tab:summary}
\end{table}

The measurements presented here, together with those given in
Ref.~\cite{PDG2012} for the $D^+$ and $D^0$ mass, and the mass
differences $M(D^+) -M(D^0)$, $M(D^{+}_s) - M(D^{+})$ 
can be used to determine a more precise value of the $D^+_s$ mass
\begin{equation}
M(D^{+}_s)  =  1968.19 \pm 0.20 \pm 0.14 \pm 0.08 \mevcc, \nonumber
\end{equation}
where the first uncertainty is the quadratic sum of the statistical
and uncorrelated systematic uncertainty, the second is due to the
momentum scale and the third due to the energy loss. This value is 
consistent with, but more precise than, that obtained from the 
fit to open charm mass data, $M(D^{+}_s) =  1968.49 \pm
0.32~\mevcc$ \cite{PDG2012}. 

%

% Do not include this in analysis note and conference reports
\section*{Acknowledgements}

\noindent We express our gratitude to our colleagues in the CERN
accelerator departments for the excellent performance of the LHC. We
thank the technical and administrative staff at the LHCb
institutes. We acknowledge support from CERN and from the national
agencies: CAPES, CNPq, FAPERJ and FINEP (Brazil); NSFC (China);
CNRS/IN2P3 and Region Auvergne (France); BMBF, DFG, HGF and MPG
(Germany); SFI (Ireland); INFN (Italy); FOM and NWO (The Netherlands);
SCSR (Poland); ANCS/IFA (Romania); MinES, Rosatom, RFBR and NRC
``Kurchatov Institute'' (Russia); MinECo, XuntaGal and GENCAT (Spain);
SNSF and SER (Switzerland); NAS Ukraine (Ukraine); STFC (United
Kingdom); NSF (USA). We also acknowledge the support received from the
ERC under FP7. The Tier1 computing centres are supported by IN2P3
(France), KIT and BMBF (Germany), INFN (Italy), NWO and SURF (The
Netherlands), PIC (Spain), GridPP (United Kingdom). We are thankful
for the computing resources put at our disposal by Yandex LLC
(Russia), as well as to the communities behind the multiple open
source software packages that we depend on.

\addcontentsline{toc}{section}{References}
\ifx\mcitethebibliography\mciteundefinedmacro
\PackageError{LHCb.bst}{mciteplus.sty has not been loaded}
{This bibstyle requires the use of the mciteplus package.}\fi
\providecommand{\href}[2]{#2}

%\bibliographystyle{LHCb}
%\bibliography{main,LHCb-PAPER,LHCb-CONF,LHCb-DP,local}

%\input{supplementary}

\end{document}